\documentclass[prd,a4paper]{revtex4}

\usepackage[pdftex]{graphicx}
\usepackage{psfrag}
\usepackage{inputenc}
\usepackage{amsmath}
\usepackage{amssymb}
\usepackage{subfigure}

\inputencoding{latin9}

\newcommand{\nc}{\newcommand}

\def\lsim{\; \raise0.3ex\hbox{$<$\kern-0.75em
      \raise-1.1ex\hbox{$\sim$}}\; }
\def\gsim{\; \raise0.3ex\hbox{$>$\kern-0.75em
      \raise-1.1\textmd{}ex\hbox{$\sim$}}\; }

\nc{\be}[1]{\begin{equation}\mbox{$\label{#1}$}}
\nc{\bea}[1]{\begin{eqnarray} \mbox{$\label{#1}$}}
\nc{\Section}[2]{\section{#2}\label{#1}}
\nc{\Bibitem}[1]{\bibitem{#1}}
\nc{\Label}[1]{\label{#1}}
\nc{\ie}{{\em i.e. }}
\nc{\eg}{{\em e.g. }}
\nc{\eea}{\end{eqnarray}}
\nc{\ee}{\end{equation}}
\nc{\w}{\omega}
\begin{document}

\title{Curvaton decay into relativistic matter}

\author{J. Sainio}
\thanks{jtksai@utu.fi}
\affiliation{Department of Physics and Astronomy,
University of Turku, FIN-20014 Turku, FINLAND}
\author{I. Vilja}
\thanks{vilja@utu.fi}
\affiliation{Department of Physics and Astronomy,
University of Turku, FIN-20014 Turku, FINLAND}
\date{\today}

\begin{abstract}
We consider an inflationary curvaton scenario, where the curvaton decays into two non-interacting relativistic fluids 
and later during the cosmological evolution one of them becomes non-relativistic, forming dark matter component of 
the universe. We study the thermic properties and the generation of non-gaussianity in this three fluid curvaton model. 
By solving the evolution of the system and using several cosmological conditions we find that the allowed parameter 
space is strongly constrained.
The naturalness of this curvaton scenario is also discussed.

\end{abstract}

\maketitle

\section{Introduction}

The most recent data gathered by WMAP \cite{Komatsu:2003fd,Spergel:2006hy,Komatsu:2008hk} is consistent with the hypothesis that the perturbations in the cosmic microwave background radiation (CMB) were generated in an era of cosmic inflation. This is generally achieved with a slowly rolling scalar field which leads to an exponential expansion of the universe and the observed anisotropy is generated by the fluctuations of this inflaton field \cite{Linde:1986,Liddle:2000}. Such a minimal scenario leads to adiabatic and gaussian perturbations in accordance with the current data.

A well motivated alternative to the simplest inflationary scenario is the curvaton mechanism \cite{Enqvist:2001zp,Lyth:2001nq,Moroi:2001ct,Assadullahi:2007uw,Bartolo:2002vf,Moroi:2002rd,Fujii:2002yx,Lyth:2002my,Sloth:2002xn,Hebecker:2002xw,Hofmann:2002gy,Moroi:2002vx,Enqvist:2002rf,Postma:2002et,Feng:2002nb,Gordon:2002gv,Dimopoulos:2002hm,Dimopoulos:2002kt,Liddle:2003zw,Dimopoulos:2003ii,Enqvist:2003mr,Lyth:2003ip,Dimopoulos:2003ss,Dimopoulos:2003az,Kasuya:2003va,Endo:2003fr,Hamaguchi:2003dc,Bartolo:2003jx,Giovannini:2003mb,McDonald:2003jk,Chun:2004gx} in which the perturbations are generated by a second scalar field, called the curvaton, which stays subdominant during inflation and the actual expansion of space is still driven by the inflaton. This allows the inflation potential to have more natural properties compared to the single field scenario \cite{Gordon:2002gv}. 
 However, the extra degrees of freedom in the system now allow for the possibility that the final state is not necessarily purely adiabatic.
Instead, the generation of an observable amount of isocurvature perturbations is a possibility that
can distinguish the curvaton scenario from the simple single field inflationary model.

Early models of curvaton cosmology usually assumed that the decay of the curvaton field
produces only radiation \cite{Enqvist:2001zp,Lyth:2001nq}. Later models included the possibility of radiation and 
matter production \cite{Gupta:2003jc,Lemoine:2006sc}. In the present paper we have expanded this line of thought to include the possibility that the curvaton decays into two non-interacting relativistic components. This makes it possible to model a cold dark matter component that is initially relativistic but turns non-relativistic as the universe expands and dilutes the energy densities. This process can lead to some interesting consequences: for example the curvature perturbations of matter are no longer conserved after curvaton decay if the decay occurs when the matter component is still relativistic.

In addition to the background evolution we will also solve numerically the first and second order perturbation
equations with different initial values and search for physically plausible initial values in
a parameter space based on the initial values of these variables. Besides studying the correctness of the thermodynamical evolution of the system we will also calculate the generated level of non-gaussianity and compare it to the limits given by WMAP 5 year data \cite{Komatsu:2008hk}. This strategy makes it possible to study the naturalness of different dark matter particle species within the curvaton decay model.


This paper is organized as follows. In section II we present the relevant quantities and equations of motion of the perturbation up to second order. In section III we present the equations of motion of the curvaton model and derive the equations of motion for the temperatures for radiation and dark matter. In section IV we present numerical results of the simulations and we end this article with discussion and conclusions in section V.

\section{Perturbations at first and second order}

The theory of second order cosmological perturbations has been studied rigorously in the recent years and a good introduction to the subject can be found {\it e.g.} in \cite{Bartolo:2004if}. We adopt their notations and conventions, and employ a spatially flat Friedmann-Robertson-Walker -background. The metric tensor up to second order is \cite{Bartolo:2004if}
\be{metric}
\begin{aligned}
g_{\mu\nu}dx^{\mu}dx^{\nu} = & -(1+2\phi^{(1)}+\phi^{(2)})dt^2+a(t)(\hat{\omega}^{(1)}_{i}+\frac{1}{2}\hat{\omega}^{(2)}_{i})dtdx^{i}\\
& +a(t)^2\Big[(1-2\psi^{(1)}-\psi^{(2)})\delta_{ij}+(\chi^{(1)}_{ij}+\frac{1}{2}\chi^{(2)}_{ij})\Big]dx^idx^j,
\end{aligned}
\ee
where $a(t)$ is the scale factor and $\phi^{(r)}$, $\hat{\omega}^{(r)}_{i}$, $\psi^{(r)}$ and $\chi^{(r)}_{ij}$ are (scalar, vector and tensor) perturbation functions 
at first ($r=1$) and second order ($r=2$). Written in this form, different gauges can be 
straightforwardly given in terms of the perturbation functions: for example the Poisson gauge is defined as $\omega^{(r)} = \chi^{(r)}_{ij} = \chi^{(r)}_{ij} = 0$ and the spatially flat one is $\psi^{(r)} = \chi^{(r)} = 0$ 

A useful set of equations can be derived from the continuity equations $T^{\mu\nu}_{i;\mu}=Q^{\nu}_{i}$, where $T^{\mu\nu}_{i}$ is the energy-momentum tensor, $Q_{i}$ describes the energy transfer between different fluids and 
semi-colon subscript ($;$) denotes the covariant derivative. 
From the continuity equations it follows that the equations determining the background evolution of individual fluids are
The background evolution equations for individual fluids follow from the continuity equations
\begin{equation} \label{eq:conti}
\dot{\rho}_{i} = -3H(1+\omega_{i})\rho_{i} + Q_{i},
\end{equation}
where $\w_{i}=P_{i}/\rho_{i}$ is the equation of state of th $i$th fluid and $\dot{} \equiv d/dt$ \ie derivative
with respect to physical time.

The evolution equations of the first order perturbed energy and pressure densities
(on large scales) are \cite{Malik:2004tf}
\be{fluidpert-1}
\dot{\delta\rho}^{(1)}_{i}+3H(\delta\rho^{(1)}_{i}+\delta P^{(1)}_{i})-3(\rho_{i}+P_{i})\dot{\psi}^{(1)} = Q_{i}\phi^{(1)} + \delta Q^{(1)}_{i}.
\ee
and at second order
\be{fluidpert-2}
\begin{aligned}
&\dot{\delta\rho}^{(2)}_{i}+3H(\delta\rho^{(2)}_{i}+\delta P^{(2)}_{i})-3(\rho_{i}+P_{i})\dot{\psi}^{(2)} - 6\dot{\psi}^{(1)}[\delta\rho_{i}+\delta P_{i}+2(\rho_{i}+P_{i})\psi^{(1)}] \\
&= Q_{i}\phi^{(2)} + \delta Q^{(2)}_{i} - Q_{i}(\phi^{(1)})^{2}+2\phi^{(1)}\delta Q^{(1)}_{i}.
\end{aligned}
\ee

These equations can be simplified further by selecting a gauge. We will be using the spatially flat gauge and the Einstein equations simplify \cite{Bartolo:2004if} in this case into conditions $\psi^{(1)}=\chi^{(1)}=0$ and $2\phi^{(1)}=-\delta\rho/\rho_{0}$ at first order and at second order the $0-0$-component of the Einstein equations gives
\begin{equation}
\phi^{(2)} = -\frac{1}{2}\frac{\delta\rho^{(2)}}{\rho_{0}}+4(\phi^{(1)})^{2}.\\
\end{equation}


\subsection{Curvature perturbations}

One way to study the evolution of perturbations is to use gauge-invariant curvature perturbations, which relate to curvature perturbations on homogeneous-density surfaces. At first order they are defined for component $i$ as
\begin{equation} \label{eq:zeta-1}
\zeta^{(1)}_{i}=-\psi^{(1)}-\frac{\delta\rho^{(1)}_{i}}{\rho_{i}'},
\end{equation}
where $'\equiv d/d(\ln(a))$. At second order the corresponding quantity is defined as
\begin{equation} \label{eq:zeta-2}
\zeta^{(2)}_{i}=-\psi^{(2)}-\frac{\delta\rho^{(2)}_{i}}{\rho_{i}'}+2\frac{\delta\rho^{(1)}_{i}{}'}{\rho_{i}'}\frac{\delta\rho^{(1)}_{i}}{\rho_{i}'}+2\frac{\delta\rho^{(1)}_{i}}{\rho_{i}'}(\psi^{(1)}{}'+2\psi^{(1)})-\Big(\frac{\delta\rho^{(1)}_{i}}{\rho_{i}'}\Big)^{2}\Big(\frac{\rho_{i}''}{\rho_{i}'}-2\Big).
\end{equation}
Note that we are here neglecting gradient terms since we are only interested in the large scale behavior of perturbations.

The equations of motion of the first order curvature perturbations can be derived from eq. (\ref{fluidpert-1}) and Einstein equations and this results in
\begin{equation} \label{eq:zeta-eq-1}
\begin{aligned}
\zeta^{(1)}_{i}{}' = & \frac{3\delta P_{int(i)}}{\rho_{i}'} - \frac{\delta Q_{int (i)}}{H\rho_{i}'} -\frac{H'}{H}\frac{Q_{i}}{\rho_{i}'}(\zeta - \zeta_{i}),\\
\end{aligned}
\end{equation}
where $\delta P_{int(i)} \equiv \delta P_{i} - p_{i}'\delta \rho_{i}/\rho_{i}'$ and $\delta Q_{int (i)} \equiv \delta Q_{i}-Q'_{i}\delta \rho_{i}/\rho'_{i}$.

At second order the corresponding equations read as \cite{Bartolo:2003bz}
\begin{equation} \label{eq:zeta-eq-2}
\begin{aligned}
\zeta^{(2)}_{i}{}' = & -\frac{1}{\rho_{i}'H}\Big[\Big(\delta Q^{(2)}_{i}-\frac{Q_{i}'}{\rho_{i}'}\delta\rho^{(2)}_{i}\Big)+Q_{i}\frac{\rho_{0}'}{2\rho_{0}}\Big(\frac{\delta\rho^{(2)}_{i}}{\rho_{i}'}-\frac{\delta\rho^{(2)}}{\rho_{0}'}\Big)\Big]-3\frac{Q_{i}}{\rho_{i}'H}\Big(\phi^{(1)}\Big)^2\\
& -2\frac{\delta Q^{(1)}_{i}\phi^{(1)}}{H\rho_{i}'}+2\Big[2-3(1+\omega_{i})\Big]\zeta^{(1)}_{i}\zeta^{(1)}_{i}{}'-2\Big[\Big(\frac{Q_{i}\phi^{(1)}}{H\rho_{i}'}+\frac{\delta Q^{(1)}_{i}}{H\rho_{i}'}\Big)\zeta^{(1)}_{i}\Big]' - \Big[\Big(\frac{Q_{i}'}{H\rho_{i}'}-\frac{1}{2}\frac{Q_{i}}{H\rho_{i}'}\frac{\rho_{0}'}{\rho_{0}}\Big)\Big(\zeta^{(1)}_{i}\Big)^{2}\Big]'.
\end{aligned}
\end{equation}

There are instances when the definition of different curvature perturbations might fail \eg when $\rho_{i}'=0$. Therefore in our numerical evaluations we have used the spatially flat gauge and evaluated the density perturbations of different components in this gauge at first and second-order. The corresponding equations of motion can be easily read from equations (\ref{fluidpert-1}) and (\ref{fluidpert-2}) by going to the spatially flat gauge $\psi^{(r)} = \chi^{(r)} = 0$.

\section{The curvaton model}

The curvaton model of the present paper consists of three fluids: curvaton, radiation, and a matter component
that is initially relativistic but becomes non-relativistic with the expansion of the universe. This happens 
roughly when the average energy of matter particles drops below its mass. This non-relativistic matter component
can thus be thought to form the dark matter of the universe.

We denote the curvaton by subscript $\sigma$, radiation by $\gamma$ and matter by $m$.
The matter and radiation components are assumed to be non-interacting with each other which means 
there are no processes that would thermalize these two components after the curvaton decay. This implies 
that the curvaton mass is higher than the Hubble rate as it starts to decay. Otherwise there would be curvaton mediated
thermalization processes that would make the matter and radiation components interacting.

In the current scenario before the curvaton decay the energy density of the universe is dominated by radiation with 
a small component of coherent curvaton field. After the curvaton decay the system consists of two fluids,
radiation and matter, with separate thermal equilibrium distributions and temperatures $T_{\gamma}$ and $T_{m}$, for the radiation and matter component respectively.
This is a practical assumption which requires that an unknown self-interactions maintains kinetic equilibrium of the matter 
component. Without this assumption the matter particle distribution should be calculated directly from the Boltzmann 
equations. This would however be quite a demanding task requiring that the coherent oscillations of
curvaton field are carefully taken into account. Thus we simplify the analysis with equilibrium distributions.
It should be noted that if there were processes maintaining equilibrium between radiation and matter
we would enter the WIMP scenario \cite{Lemoine:2006sc} or we would have $m\sim T_{eq}\leq 1$ eV and the matter component would begin to dominate
cosmic evolution immediately after becoming non-relativistic. This is hardly a plausible cosmological scenario for the dark matter \cite{Bertone:2004pz}.


\subsection{Evolution equations}

For the radiation we have $\omega_{\gamma}=1/3$ but the equation of state of the dark matter component depends on the temperature of the matter component, $T_{m}$. Initially the matter is relativistic and $\omega_{m}=1/3$ but as the universe cools down the matter turns into non-relativistic and $\omega_{m} \rightarrow 0$. Because of this $T_{m}$ has time dependence that will be derived in the next subsection. This equation naturally needs to be incorporated into the equations of motion presented in this subsection.
The interaction terms are \cite{Gupta:2003jc}
\begin{equation} \label{eq:interact}
\begin{aligned}
Q_{\sigma} & = -\Gamma_{\gamma}\rho_{\sigma} - \Gamma_{m}\rho_{\sigma}\\
Q_{\gamma} & = \Gamma_{\gamma}\rho_{\sigma}\\
Q_{m} & = \Gamma_{m}\rho_{\sigma},\\
\end{aligned}
\end{equation}
where $\Gamma_{i}$ denote the strengths of the interactions.

Background equations (\ref{eq:conti}) can be written in terms of fractional densities $\Omega_{i} \equiv \rho_{i}/\rho$ for which the equations of motion are:
\begin{equation} \label{eq:bgr}
\begin{aligned}
\Omega_{\sigma}' & =\Omega_{\sigma}(\Omega_{\gamma} + 3\omega_{m}\Omega_{m})+\frac{Q_{\sigma}}{H\rho} ,\\
\Omega_{\gamma}' & =\Omega_{\gamma}(\Omega_{\gamma} + 3\omega_{m}\Omega_{m} -1 )+\frac{Q_{\gamma}}{H\rho} ,\\
\Omega_{m}' & =\Omega_{m}\Big( \Omega_{\gamma} + 3\omega_{m}(\Omega_{m}-1) \Big)+\frac{Q_{m}}{H\rho} ,\\
\Big(\frac{1}{H}\Big)' & = \Big(1+\frac{1}{3}\Omega_{\gamma}\Big)\Big(\frac{1}{H}\Big).\\
\end{aligned}
\end{equation}
From the definition of $\Omega_{i}$ it can be easily seen that $\Omega_{\sigma}+\Omega_{\gamma}+\Omega_{m}=1$, which means that one of the $\Omega_{i}$ equations is redundant.

From eq. (\ref{fluidpert-1}) we can read the equations of motion for the first order density perturbations
\begin{equation} \label{eq:flat-rho1}
\begin{aligned}
\delta\rho^{(1)}_{\sigma}{}' & = -3\delta\rho^{(1)}_{\sigma} - \frac{Q_{\sigma}}{H}\frac{\delta\rho^{(1)}}{2\rho} + \frac{\delta Q^{(1)}_{\sigma}}{H},\\
\delta\rho^{(1)}_{\gamma}{}' & = -4\delta\rho^{(1)}_{\gamma} - \frac{Q_{\gamma}}{H}\frac{\delta\rho^{(1)}}{2\rho} + \frac{\delta Q^{(1)}_{\gamma}}{H},\\
\delta\rho^{(1)}_{m}{}' & = -3(1+\omega_{m})\delta\rho^{(1)}_{m} - \frac{Q_{m}}{H}\frac{\delta\rho^{(1)}}{2\rho} + \frac{\delta Q^{(1)}_{m}}{H}\\
\end{aligned}
\end{equation}
in the spatially flat gauge. At second order the corresponding equations in the flat gauge are
\begin{equation} \label{eq:flat-rho2}
\begin{aligned}
\delta\rho^{(2)}_{\sigma}{}' & = -3\delta\rho^{(2)}_{\sigma} - \frac{Q_{\sigma}}{2H\rho}\Bigg[\delta\rho^{(2)} -\frac{3}{2}\frac{\Big(\delta\rho^{(1)}\Big)^2}{\rho}\Bigg] + \frac{\delta Q^{(2)}_{\sigma}}{H} - \frac{\delta\rho^{(1)}\delta Q^{(1)}_{\sigma}}{H\rho},\\
\delta\rho^{(2)}_{\gamma}{}' & = -4\delta\rho^{(2)}_{\gamma} - \frac{Q_{\gamma}}{2H\rho}\Bigg[\delta\rho^{(2)} -\frac{3}{2}\frac{\Big(\delta\rho^{(1)}\Big)^2}{\rho}\Bigg] + \frac{\delta Q^{(2)}_{\gamma}}{H} - \frac{\delta\rho^{(1)}\delta Q^{(1)}_{\gamma}}{H\rho},\\
\delta\rho^{(2)}_{m}{}' & = -3(1+\omega_{m})\delta\rho^{(2)}_{m} - \frac{Q_{m}}{2H\rho}\Bigg[\delta\rho^{(2)} -\frac{3}{2}\frac{\Big(\delta\rho^{(1)}\Big)^2}{\rho}\Bigg] + \frac{\delta Q^{(2)}_{m}}{H} - \frac{\delta\rho^{(1)}\delta Q^{(1)}_{m}}{H\rho}.\\
\end{aligned}
\end{equation}

We also need the gauge invariant curvature perturbations $\zeta^{(i)}_{j}$ when calculating the non-gaussianity parameter $f_{NL}$. In the spatially flat gauge these are
\begin{equation} \label{eq:zetas-flat}
\begin{aligned}
\zeta^{(1)}_{i} =&-\frac{\delta\rho^{(1)}_i}{\rho_{i}'},\\
\zeta^{(2)}_{i} =&-\frac{\delta\rho^{(2)}_i}{\rho_{i}'}+\Big[2-3(1+\omega_{i})\Big]\Big(\zeta^{(1)}_{i}\Big)^2 - 2\Big[\frac{Q_{i}\phi^{(1)}}{H\rho_{i}'}+\frac{\delta Q^{(1)}_{i}}{H\rho_{i}'}\Big]\zeta^{(1)}_i\\
&-\Big[\frac{Q_i'}{\rho_i'H}-\frac{1}{2}\frac{Q_i \rho'}{\rho_i'H\rho}\Big]\Big(\zeta^{(1)}_{i}\Big)^2.
\end{aligned}
\end{equation}

The set of equations (\ref{eq:bgr}), (\ref{eq:flat-rho1}) and (\ref{eq:flat-rho2}) can now be evaluated numerically once the initial values have been set. We have chosen the system to be initially radiation dominated and non-adiabatic at first and second-order. The initial values for the three fluid curvaton decay were derived in \cite{Multamaki:2008yv} and they are given by
\begin{equation} \label{eq:zetas-flat-init}
\begin{aligned}
\zeta^{(1)}_{\sigma,\textrm{in}} =& 1,\\
\zeta^{(2)}_{\sigma,\textrm{in}}\simeq & 1/2\\
\end{aligned}
\end{equation}
and the latter approximation is valid when $\Gamma_{\sigma}\ll H_0$.
The other perturbations are initially set to be zero
\ie $\zeta^{(1)}_{\gamma,\textrm{in}}=\zeta^{(1)}_{m,\textrm{in}}=\zeta^{(2)}_{\gamma,\textrm{in}}=\zeta^{(2)}_{m,\textrm{in}}=0$ since we have assumed that only the curvaton field has initial perturbations.

\subsection{Thermodynamics of the system}

Our curvaton models starts in a radiation dominated universe with initial radiation and dark matter temperature set to $T_{\gamma,\textrm{in}}$ and $T_{m,\textrm{in}}$ respectively. We will assume that the dark matter particles have a mass $m_{m}$ and by setting the initial matter temperature $T_{m,\textrm{in}}$ we can change the initial equation of state $\omega_{m}$ of the dark matter. Different temperatures for radiation and dark matter mean that there are no interactions between them and the temperature $T_{m}$ is due to kinetic equilibrium of the dark matter particles.

We will start by deriving the equations of motion for the variables that govern the evolution of the system
\ie the temperatures $\beta_{\gamma}$ and $\beta_{m}$ where $\beta = T^{-1}$.
The energy density of dark matter particles can be naturally written as
\begin{equation} \label{eq:en-dens}
\begin{aligned}
\rho_{m} & = \int \frac{dp^3}{(2\pi)^3} E f_{m} = C \int_{0}^{\infty} dp p^2 E f_{m}(\beta_{m}(E-\mu))
\end{aligned}
\end{equation}
where we have made the ansatz $f_{m}=f_{m}(\beta_{\textrm{m}}(E-\mu))$ for the distribution function and used the notation $C = 1/(2\pi^2)$.
The pressure and number density are defined similarly as
\begin{equation} \label{eq:pres}
P_{\textrm{m}} = \int \frac{dp^3}{(2\pi)^3} \frac{p^2}{3E} f_{m}(\beta_{m}(E-\mu))
\end{equation}
and
\begin{equation} \label{eq:nden}
n_{\textrm{m}} = \int \frac{dp^3}{(2\pi)^3} f_{m}(\beta_{m}(E-\mu)).
\end{equation}

Starting from the Boltzmann equation we could now derive the equations of motion for the energy density, pressure and number density. However since Boltzmann equation is equivalent to the continuity equations (\ref{eq:conti}) we will only use the latter.
Therefore by differentiating equation (\ref{eq:en-dens}) with respect to time we get
\begin{equation} \label{eq:en-dens-dot1}
\begin{aligned}
\dot{\rho}_{m} & = C \int_{0}^{\infty} dp p^2 E f'_{m}\dot{\beta}_{m}(E-\mu) = 
\dot{\beta}_{m} \Big[ C  \int_{0}^{\infty} dp p \frac{E^2(E - \mu)}{{\beta}_{m}}\frac{df_{m}}{dp}\Big]\\
& = - \frac{\dot{\beta}_{m}}{\beta_{m}} \Big[ C \int_{0}^{\infty} dp \frac{d}{dp}\Big(p(E^3-\mu E^2)\Big) f_{m} \Big]
 = - \frac{\dot{\beta}_{m}}{\beta_{m}} \Big[ C \int_{0}^{\infty} dp \Big(E^3-\mu E^2 + 3p^2 E - 2\mu p^2\Big) f_{m} \Big]\\
& = - \frac{\dot{\beta}_{m}}{\beta_{m}} \Big[ C \int_{0}^{\infty} dp \Big(E(p^2+m^2)-\mu(p^2+m^2)\Big) f_{m} + 3 \rho_{m} - 2\mu n_{m} \Big]\\
& = - \frac{\dot{\beta}_{m}}{\beta_{m}} \Big[ C \int_{0}^{\infty} dp \Big(m^2(E-\mu)\Big) f_{m} + 4 \rho_{m} - 3\mu n_{m} \Big], \\
& = - \frac{\dot{\beta}_{m}}{\beta_{m}} \Big[ m^2\Big\langle \frac{E-\mu}{p^2}\Big \rangle + 4 \rho_{m} - 3\mu n_{m} \Big], \\
\end{aligned}
\end{equation}
where prime now means derivative with respect to the argument of the distribution function and we have also made partial integration in the third equality where the assumption $\mu = m_{m}$ valid for the dark matter particles was used. We can now combine this equation with the continuity equation (\ref{eq:conti}) and solve the equation for $\dot{\beta}_{m}$. After some algebra we get the result
\begin{equation} \label{eq:beta}
\begin{aligned}
\frac{\dot{\beta}_{m}}{\beta_{m}} & = \frac{3H(\rho_{m}+P_{m})-\Gamma_{\sigma}\rho_{\sigma}}{4\rho_{m}-3\mu n_{m} + m^2\Big\langle \frac{E-\mu}{p^2}\Big \rangle}
= \frac{3H(1+\omega_{m})-\Gamma_{\sigma}\Omega_{\sigma}/\Omega_{m}}{4-3\chi_{m} + \Big\langle \frac{E-\mu}{p^2}\Big \rangle m^2/\rho_{m}}.
\end{aligned}
\end{equation}
Note that in the last equality we have also used a new variable $\chi_{\textrm{m}} = mn_{\textrm{m}}/\rho_{\textrm{m}}$ which is close to zero for relativistic matter and one for non-relativistic case.

The temperature in the early universe is generally very high which means that both Bose-Einstein and Fermi-Dirac statistics can be well approximated with the Maxwell-Boltzmann distribution. The distribution of the the dark matter particles can be therefore written as \ie $f_{m}=e^{-\beta_{\textrm{m}}(E-\mu)}$. Note that this has also pragmatic reasons that will be evident in the numerical calculations.
By substituting this distribution in the definition of energy density and after some integral transformations the result can be written in terms modified Bessel functions $K_{n}(x)$:
\begin{equation} \label{eq:en-dens1}
\begin{aligned}
\rho_{\textrm{m}} & = \int \frac{dp^3}{(2\pi)^3} E f_{m} = C \int_{0}^{\infty} dp p^2 E e^{-\beta_{\textrm{m}}(E-\mu)} = 
C \int_{0}^{\infty} dp p^2 \sqrt{p^2+m^2} e^{-\beta_{\textrm{m}}(\sqrt{p^2+m^2} - \mu)} \\
& = C e^{\beta_{\textrm{m}} \mu } \int_{m}^{\infty} dE E^{2} \sqrt{E^2-m^2} e^{-\beta_{\textrm{m}}E} =
C e^{\beta_{\textrm{m}} \mu } m^4 \int_{1}^{\infty} dx x^{2} \sqrt{x^2 - 1} e^{-\beta_{\textrm{m}}mx} \\
& = C e^{\beta_{\textrm{m}} \mu } \frac{m^2}{\beta_{\textrm{m}}^{2}} \Big( m\beta_{\textrm{m}} K_{1}(m\beta_{\textrm{m}}) + 3 K_{2}(m\beta_{\textrm{m}})\Big)
\end{aligned}
\end{equation}
where in the last equality we have used modified Bessel functions of the second kind $K_{2}(x)$.
The pressure and number density can be written similarly in terms of Bessel functions as
\begin{equation} \label{eq:pres1}
P_{\textrm{m}} = \int \frac{dp^3}{(2\pi)^3} \frac{p^2}{3E} f_{m} = C e^{\beta_{\textrm{m}} \mu } \frac{m^2}{\beta_{\textrm{m}}^{2}} K_{2}(m\beta_{\textrm{m}})
\end{equation}
and
\begin{equation} \label{eq:nden1}
n_{\textrm{m}} = \int \frac{dp^3}{(2\pi)^3} f_{m} = C e^{\beta_{\textrm{m}} \mu } \frac{m}{\beta_{\textrm{m}}} K_{2}(m\beta_{\textrm{m}}).
\end{equation}

The equation of state of the dark matter particles is defined as the ratio of pressure to density and by using equations (\ref{eq:en-dens1}) and (\ref{eq:pres}) this simplifies to
\begin{equation} \label{eq:omega}
\omega_{m} = \frac{K_{2}(m\beta_{m})}{m\beta_{m} K_{1}(m\beta_{m}) + 3 K_{2}(m\beta_{m})}.
\end{equation}
We can also write the $\chi_{\textrm{m}}$ function in terms of Bessel functions as
\begin{equation} \label{eq:chi}
\chi_{m} = \frac{m\beta_{m}K_{2}(m\beta_{m})}{m\beta_{m} K_{1}(m\beta_{m}) + 3 K_{2}(m\beta_{m})}.
\end{equation}
The final term to be integrated is $\langle (E-\mu)/p^2 \rangle$ which equals now
\begin{equation} \label{eq:en-E-mu}
\Big\langle \frac{E-\mu}{p^2} \Big\rangle = \frac{m e^{\mu \beta_{m}}}{\beta_{m}}\Big(m\beta_{m}K_{0}(m\beta_{m}) + K_{1}(m\beta_{m})-\frac{\mu}{m}m^2K_{1}(m\beta_{m})\Big).
\end{equation}

\begin{figure}[tbh]
\includegraphics[width=0.45\columnwidth]{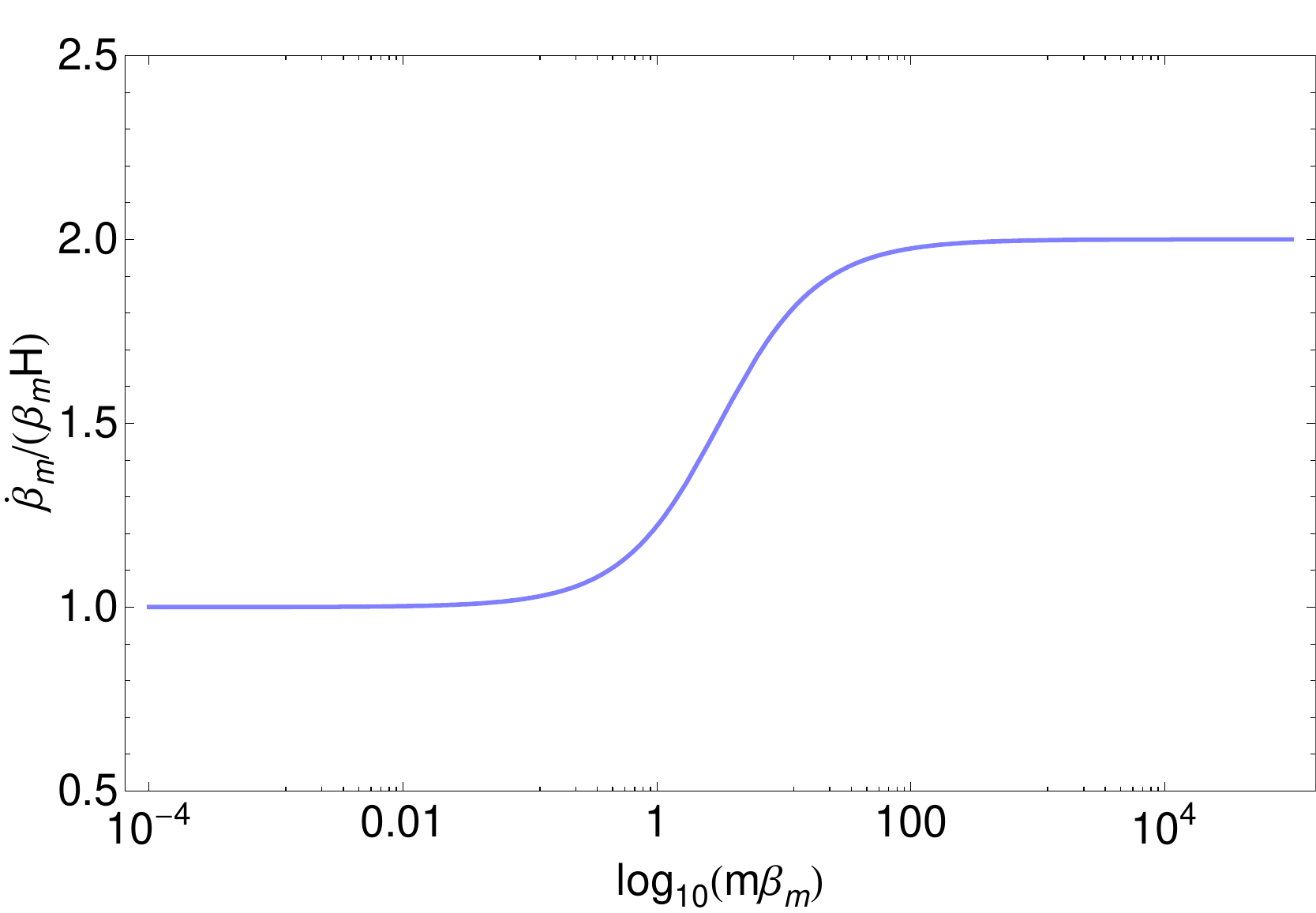}
\caption{Plot of $\dot{\beta_{m}}/(\beta_{m}H)$ (equation (\ref{eq:beta-2})) when there is no curvaton decay into matter as a function of the logarithm of $m\beta_{m}$. When $m\beta_{m} \ll 1$ or equivalently $T_{m} \gg m$ $\dot{\beta}_{m}/\beta_{m} \propto H$ which corresponds to the usual relativistic limit $T_{m} \propto a^{-1}$. In the non-relativistic case $m\beta_{m} \gg 1$ (or equivalently $T_{m} \ll m$) $\dot{\beta}_{m}/\beta_{m} \propto 2 H$ which leads to the usual non-relativistic evolution $T_{m} \propto a^{-2	}$. Equation (\ref{eq:beta-2}) thus clearly leads to physically sound evolution for the temperature.}
\label{fig1}
\end{figure}

These expressions can now be substituted in the temperature equation (\ref{eq:beta}) and use some of the identities of Bessel functions to get the final result
\begin{equation} \label{eq:beta-2}
\begin{aligned}
\frac{\dot{\beta}_{m}}{\beta_{m}} &
= \frac{3H \Big(m \beta_{m} K_{1}(m\beta_{m}) + 4 K_{2}(m\beta_{m})\Big)-\Gamma_{\sigma}\Omega_{\sigma}/\Omega_{m} \Big(m \beta_{m} K_{1}(m\beta_{m}) + 3 K_{2}(m\beta_{m})\Big)}{m \beta_{m} \Big(m \beta_{m} K_{0}(m \beta_{m}) + (5-\mu \beta_{m})K_{1}(m\beta_{m})\Big) + 3(4 - \mu \beta_{m}) K_{2}(m\beta_{m})}\\
\end{aligned}
\end{equation}
which determines the evolution of the temperature of dark matter particles.
We have plotted this function in terms of Hubble parameter, $H$, in figure \ref{fig1} when $\Gamma_{m}=0$ \ie when there is no curvaton decay into matter. The figure clearly shows that the equations leads to the proper relativistic, $T_{m} \propto a^{-1}$, and non-relativistic, $T_{m} \propto a^{-2}$, limits for the temperature evolution.

We now have a complete set of equations that determine the evolution of the system: namely equations (\ref{eq:bgr}), (\ref{eq:flat-rho1}), (\ref{eq:flat-rho2}), (\ref{eq:omega}), (\ref{eq:chi}) and (\ref{eq:beta-2}) that we can solve for the evolution of the background and the perturbations at first and second order. Note that when solving these equations we have used $N=\ln(a)$ as the time variable.

\section{Numerical results}

We have solved the evolution of the system (\ref{eq:bgr}), (\ref{eq:flat-rho1}), (\ref{eq:flat-rho2}), (\ref{eq:omega}), (\ref{eq:chi}) and (\ref{eq:beta-2}) with a range of different initial values and classified them to physical and non-physical regions \ie initial values that could have produced the universe we observe. We used a similar strategy previously in \cite{Multamaki:2007hv} for a three fluid curvaton decay where the curvaton decays into radiation and non-relativistic matter. In order to label different points into physical and non-physical regions we have made a range of tests that system needs to pass in order to be classified as physically viable. These are:

1) We check that the curvaton starts to decay before nucleosynthesis. If this condition is not met the universe might be radiation dominated at nucleosynthesis but as the curvaton component behaves like a non-relativistic fluid it will eventually start to dominate the universe and therefore make the nucleosynthesis non-physical.

2) The universe should be radiation dominated at nucleosynthesis temperature which has been set to be $\sim 10^5$ eV \cite{Liddle:2000}. In terms of fractional densities this corresponds to \cite{Multamaki:2007hv}
\begin{equation} \label{eq:t1}
\frac{\Omega_{\gamma}}{\Omega_{\sigma}+\Omega_{m}}\Big|_{\textrm{nuc}} \simeq \frac{T_{\textrm{nuc}}}{T_{\textrm{eq}}} 
\geq 10^{5}.
\end{equation}
A different choice would naturally lead to different allowed regions but as was noted in \cite{Multamaki:2007hv} the changes are not drastic.

3) The radiation matter equality should happen at $\sim 0.706$ eV \cite{Amsler:2008zzb}. This turns out to be quite strict limit in terms of the resolution of the simulations we have computed. We have therefore easened this to a range of possible equality temperatures: $0.6 \textrm{ eV } < T_{eq} < 0.8 \textrm{ eV }$. This limitation could be circumvented by increasing the number of different initial values (and thereby increasing the resolution of the simulation) but this quickly leads to very long simulation times. This way we instead get a region which includes the initial values corresponding to the more strict equality temperature.

4) The $f_{NL}$ should be within the limits $-9 < f_{NL} < 111$ given by WMAP 5 year data \cite{Komatsu:2008hk}. The equation for the non-gaussianity in the three fluid curvaton decay was derived in \cite{Multamaki:2008yv} and we will only cite the result
\begin{equation} \label{eq:fNL}
f_{NL}=\frac{25(q_2-q_1^2)-60q_1r_1+96r_1^2-30r_2}{6r_1(6r_1-5q_1)},
\end{equation}
where
\begin{equation} \label{eq:rq}
\begin{aligned}
r_{1} & = \frac{\zeta^{(1)}_m|_m}{\zeta^{(1)}_{\sigma,\textrm{in}}}, \quad q_{1}=\frac{\zeta^{(1)}_{\gamma}|_m}{\zeta^{(1)}_{\sigma,\textrm{in}}}\\
r_{2} & =\frac{\zeta^{(2)}_m|_m}{(\zeta^{(1)}_{\sigma,\textrm{in}})^2}, \quad q_{2}=\frac{\zeta^{(2)}_{\gamma}|_m}{(\zeta^{(1)}_{\sigma,\textrm{in}})^2}\\
\end{aligned}
\end{equation}
and where the different numerators are evaluated at the time of decoupling when the universe is matter dominated and hence $\zeta^{(i)}\simeq\zeta^{(i)}_{m}$. The curvaton initial perturbation values were given in equations (\ref{eq:zetas-flat-init}). Note that the system is adiabatic if $q_{1}=r_{1}$ and $q_{2}=r_{2}$. The physical meaning of $r_i$ and $q_i$ is that they tell how efficiently the initial curvature perturbations are converted into the matter and radiation components.

Before the simulations can be run different initial values have to be set: namely 
the initial temperatures of radiation $T_{\gamma,\textrm{in}}$ and the initial temperature $T_{m,\textrm{in}}$ and particle mass $m_{m}$ for the dark matter, which also define the corresponding energy densities, and the initial fractional density of the curvaton. The initial Hubble parameter value can be calculated from these variables and the interaction strengths $\Gamma_{i}$ vary from $10^{-30}$ to $10^{5}$ thereby defining the resolution of the simulations.

In the usual curvaton hypothesis the curvaton field is usually assumed to be subdominant after inflation and the relativistic decay products of inflaton dominate the energy density. The initial temperature of the system is therefore closely related to the reheating temperature. The simulations have been run with a different initial temperatures corresponding to values $T_{\gamma,\textrm{in}}=10^{14}, 10^{16}, 10^{18} \textrm{ and } 10^{20}$ eV. 

Since the dark matter particles are produced by the curvaton decay their energy density is assumed to be initially small relative to radiation.
Note that when starting the simulations equations (\ref{eq:beta}) and (\ref{eq:beta-2}) are clearly ill defined when $\Omega_{m} \rightarrow 0$. We have therefore set initially $\Omega_{m}\neq0$ which can be interpreted that when the simulation starts, the decay of the curvaton field has already started. The initial dark matter temperature, $T_{m}$, has been set to be two orders of magnitude smaller in value than the radiation temperature. Other values might lead to numerical problems during the evaluations and this selection leads to very small initial dark matter fractional densities, \ie $\Omega_{m,\textrm{in}} \ll 1$.


The amount of curvaton in the system is also a free parameter. However since we are following the usual curvaton hypothesis initially the curvaton is subdominant. This translates to values $\Omega_{\sigma,\textrm{in}} = 10^{i}$, $i = -4,-3,-2 \textrm{ and } -1$ in the different simulations.

Since we have made weak assumptions on the type of the dark matter, the mass $m_{m}$ can have a wide range of different values (A review of different dark matter particle models can be found for example in \cite{Bertone:2004pz}). The only limitations are that initially dark matter is relativistic \ie $T_{m} \gg m$ and that it does not interact with radiation. This makes it possible to consider a wide range of different dark matter particle species and to study their naturalness in the curvaton scenario.

\begin{figure}[h]
\subfigure[]{\label{fig2a}\includegraphics[width=0.45\columnwidth]{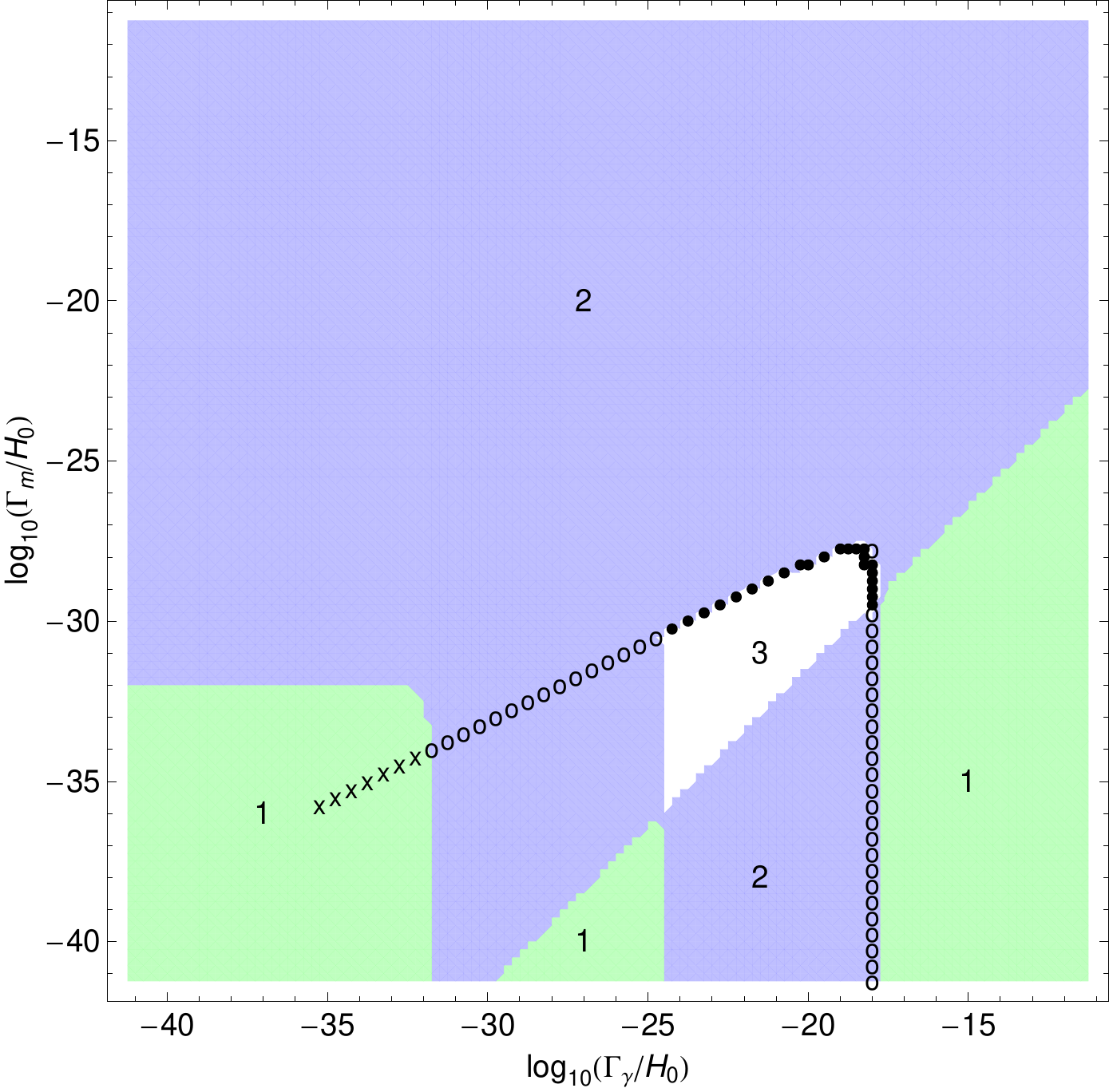}}
\quad
\subfigure[]{\label{fig2b}\includegraphics[width=0.45\columnwidth]{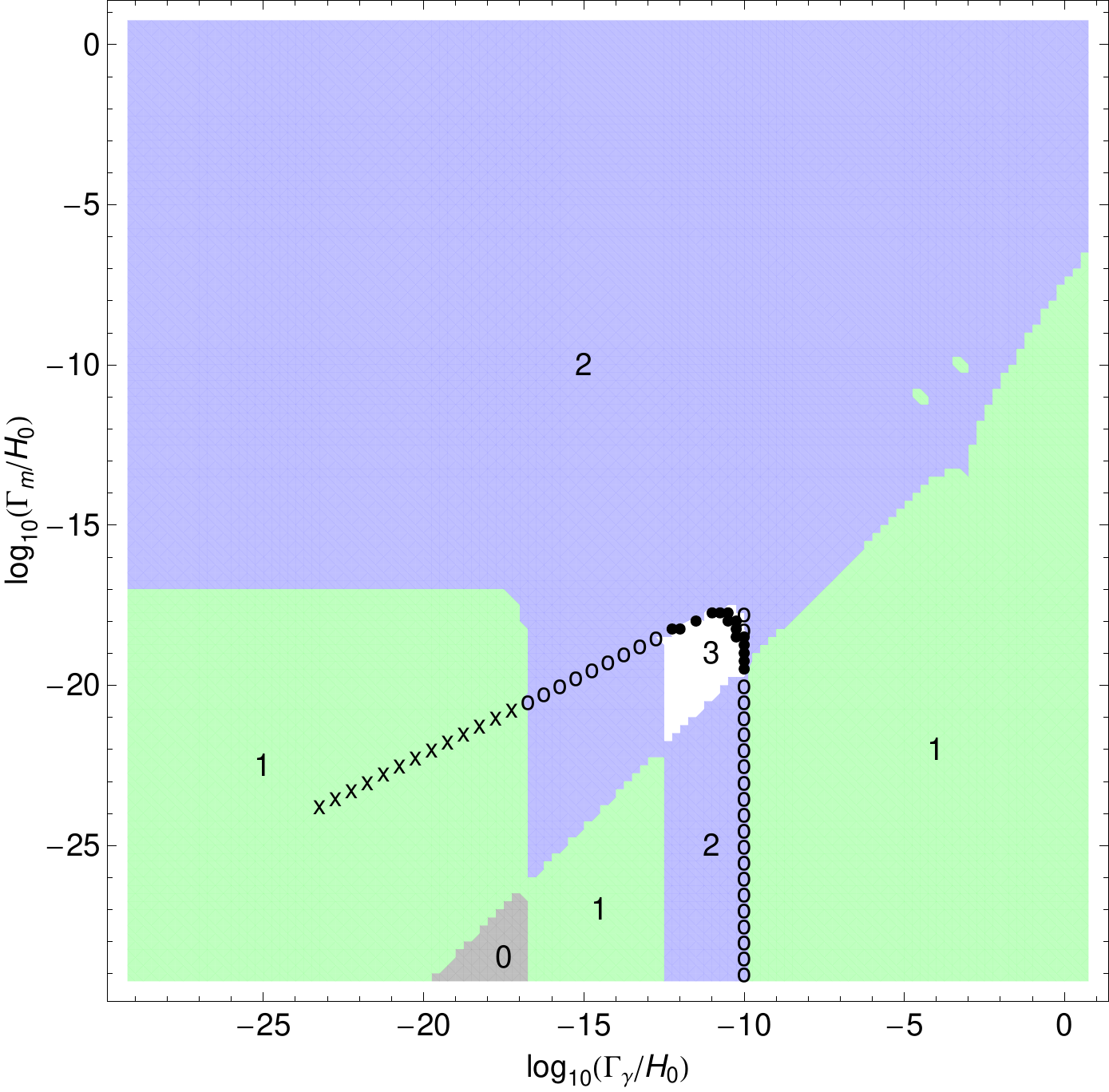}}
\caption{Contour plot of the system when (a) $T_{\gamma,\textrm{in}}=10^{18}$ eV, 
$T_{m,\textrm{in}}=10^{16}$ eV, $m_{m}=10^{13}$ eV and $\Omega_{\sigma,\textrm{in}}=0.1$ and
(b) $T_{\gamma,\textrm{in}}=10^{12}$ eV, $T_{m,\textrm{in}}=10^{10}$ eV, $m_{m}=10^{9}$ eV 
and $\Omega_{\sigma,\textrm{in}}=10^{-1}$. The plot shows the physicality tests of the 
system for different values of $\log_{10}(\Gamma_{\gamma}/H_{0})$ and 
$\log_{10}(\Gamma_{m}/H_{0})$. The numbers in different regions indicate the number of passed 
tests. Black points at the edge of white region are the points that passed all the tests, while
crosses and circles mark points which passed the matter-radiation equality test.  
Different initial values lead to qualitatively similar results.}
\end{figure}

Typical results of the simulations are shown in figures \ref{fig2a} and \ref{fig2b}. The $\log_{10}(\Gamma_{\gamma}/H_{0})$ and $\log_{10}(\Gamma_{m}/H_{0})$ plane is divided into several separate areas:
in the lower left corner the decay widths are too small and the curvaton field dominates the evolution of the system even at $T_{eq}=0.7$ eV.
The triangular area at the lower right corner is caused by the non-gaussianity limit. Inside this region $\Gamma_{\gamma}$ is much larger than $\Gamma_{m}$ and most of the curvaton decays into radiation. In terms of non-gaussianity parameter, eq. (\ref{eq:fNL}), this means that $r_{1}$ is very small and the generated non-gaussianity is very large and therefore labeled as unphysical.
The nucleosynthesis limit, eq. (\ref{eq:t1}), is satisfied inside the trapezoid shaped region between values $\log_{10}(\Gamma_{\gamma}/H_{0})=-25$ and $\log_{10}(\Gamma_{\gamma}/H_{0})=-18$ in figure \ref{fig2a}.
Points that also satisfy the radiation matter equality limit, $0.6 \textrm{ eV } < T_{eq} < 0.8 \textrm{ eV }$, are around the outer edge of this nucleosynthesis region. Despite the easened limitations for the equality temperature, this clearly gives only a very narrow region of plausible initial values.

In terms of the naturalness of the curvaton scenario, the decay widths of the curvaton are limited to be very small if the system is compelled to  satisfy the previously presented physicality tests. Smaller initial temperatures and dark matter particle mass allow larger values but even in this case the decay widths require fine-tuning.

\section{Discussion and conclusions}

We have studied the behavior of a three fluid curvaton decay model when the possibility of more complex thermodynamics of the cold dark matter decay product have been also taken into account. The assumptions we have made are relatively loose: the decay products of the curvaton were assumed to be non-interacting. This is quite a natural, if the other fluid is interpreted as conventional radiation and the other as dark matter.
The system was simulated for a wide range of different initial values including initial temperature, dark matter particle mass and curvaton decay widths. This made it possible to study the naturalness of the curvaton model in this model.

We found that the generated non-gaussianity in the system can be considerable but these large values are reached with initial values that do not lead to physically sound evolution for the system. The non-gaussianity is therefore expected to be of the same order as in the simple inflaton models \cite{Bartolo:2004if}.

In terms of the dark matter particle mass, $m_{m}$, the curvaton model leads to physically sound evolution with a range of different values but not without fine tuning in the decay widths. Lower initial temperatures and mass $m_{m}$ easen this issue but not considerably. We therefore conclude that the current curvaton decay model is not satisfactory from the point of view of naturalness.

According to our analysis,  a related, but more viable and physically sound model could be the curvaton decay into weakly interacting massive particles (WIMPs) \cite{Lemoine:2006sc}. Interactions between cold dark matter and photons would keep them at the same temperature in the early universe until they freeze-out as the temperature drops below their mass.
The same strategy we used here could be applied also to this scenario to study the naturalness of the curvaton WIMP scenario. This is however beyond the scope of this paper and is left as a future work.

\subsection*{Acknowledgments}
JS is supported by the Academy of Finland project no. 8111953.

\newpage


\end{document}